\documentclass[12pt]{article} 

\usepackage{epic,eepic,framed,epsf,latexsym,amsmath,amssymb,amscd,mathrsfs,slashbox }
\usepackage{graphicx}

\newtheorem{theorem}{Theorem}[section]
\newtheorem{lemma}[theorem]{Lemma}

\newtheorem{note}[theorem]{Note}
\newtheorem{proposition}[theorem]{Proposition}
\newtheorem{open question}[theorem]{Open question}
\newtheorem{corollary}[theorem]{Corollary}
\newtheorem{conjecture}[theorem]{Conjecture}
\newtheorem{definition}[theorem]{Definition}

\newcommand{\ignore}[1]{}

\newcommand{\proof}{{\par\noindent {\bf Proof}\space\space}}
\newcommand{\proofbox}{\begin{flushright}$\Box$\end{flushright}}

\title{Tight products and Graph Expansion}

\author{
  Amit Daniely\thanks{Department of Mathematics,
    The Hebrew University, Jerusalem, Israel,
    {\tt amit.daniely@mail.huji.ac.il}}
\and
  Nathan Linial\thanks{Department of Computer Science,
    The Hebrew University, Jerusalem, Israel,
    {\tt nati@cs.huji.ac.il}}
}

\date{January 20, 2010}

\begin{document} 
\maketitle
\begin{abstract}
In this paper we study a new product of graphs called
{\em tight product}. A graph $H$ is said to be a tight product of
two (undirected multi) graphs $G_1$ and $G_2$,
if $V(H)=V(G_1)\times V(G_2)$ and both projection maps $V(H)\rightarrow V(G_1)$
and $V(H)\rightarrow V(G_2)$ are covering maps. It is not a priori
clear when two given graphs have a tight product (in fact, it
is $NP$-hard to decide). We investigate the conditions under which this
is possible. This perspective yields a new
characterization of class-1 $(2k+1)$-regular graphs.
We also obtain a new model of random $d$-regular graphs whose
second eigenvalue is almost surely at most $O(d^{3/4})$.
This construction resembles random graph lifts,
but requires fewer random bits.
\end{abstract}

\section{Introduction and Background}
\subsection{Notations and conventions}
All the graphs in this paper are undirected, possibly with
multiple edges and
self-loops. We denote the group of permutations on a set $V$
by $S_V$. A {\em $2$-factor} in a graph $G$ is a spanning subgraph
that is the disjoint union of cycles. A {\em $2$-factorization}
is a partitioning of $E(G)$ into $2$-factors. It is well known
\cite{Diestel} that every $2d$-regular graph $G$ has a $2$-factorization.
In other words, every $2d$-regular graph $G$ on vertex set $V$ can be
constructed as follows: There are $d$ permutations $\sigma_1,...,\sigma_d \in S_V$,
such that $E(G)=\{v \sigma_i(v) | v\in V, i=1,\ldots,d\}$.
We denote such a graph by $G(\sigma_1,...,\sigma_d)$. In calculating vertex degrees, multiple
edges are counted by multiplicity and by convention, the
loop corresponding to $\sigma_i(v)=v$ is also counted twice. The neighbor set of
vertex $v$ in a graph $G$ is denoted by $N_G(v)$ (or just $N(v)$). Note that $N_G(v)$ is also a multiset.

By $\vec{E}(G)$ we denote the set of {\em ordered} pairs $\overrightarrow{v_1v_2}$ such that $v_1v_2\in E(G)$.
\subsection{Expanders and Ramanujan graphs}
Let $G=(V,E)$ be an $n$-vertex $d$-regular graph. We denote the eigenvalues of its adjacency matrix by 
$d=\lambda_1 \ge \ldots \ge \lambda_n \ge -d$. We say that $G$ is
an {\em $(n,d,\lambda$) graph} if $|\lambda_i| \le \lambda$ for every $i=2,...,n$.
We recall some basic facts about expander graphs and refer the reader to~\cite{HLW06}
for a recent survey on expander graphs and the rich theory around them.
The Alon-Boppana~\cite{alon} bound states that
$\lambda_2 \ge 2\sqrt{d-1}-o_n(1)$. If
$\lambda_2 \le 2 \sqrt{d-1}$ we say that $G$ is a {\em Ramanujan} graph.
It is known (\cite{LPS88, Mar88, Mor94}) that if $d-1$ is a prime power, then there
exist infinitely many $d$-regular Ramanujan graphs (with explicit constructions).
For other values of $d$ it is still unknown whether arbitrarily large $d$-regular Ramanujan graphs exist.
A major result due to Friedman~\cite{Fri} is that for every $d \ge 3$
and every $\epsilon > 0$ almost every $d$-regular graph satisfies $\lambda_2 \le 2 \sqrt{d-1}+\epsilon$.

\subsection{Lifts of graphs}\label{sub-se:lifts}
\begin{definition}
We say that a graph $H$ is a {\em lift} of {\em a base graph} $G$ (or that $H$ {\em covers} $G$) if there
is a map $\phi:V(H)\rightarrow V(G)$ (a {\em covering map}) such that for every $v\in V(H)$,
$\phi$ maps $N_H(v)$ one to one and onto $N_G(\phi(v))$.
For every $v \in V(G)$ the set $\phi^{-1}(v)$ is called the {\em fiber} of $v$. 
Similarly, for $e \in E(G)$, we say that $\phi^{-1}(e)$ is the fiber of $e$.
(We remind the reader that multiple edges and loops are allowed and the definitions here
should be modified accordingly).
\end{definition}
We recall some basic facts on lifts of graphs and refer the reader to~\cite{AL02,AL06,ALM02,LR05}
for a more thorough account and some recent work on the subject.
\begin{proposition}\label{lifts basic properties}
Let $\phi:V(H)\rightarrow V(G)$ be a covering map between two finite graphs.
\begin{enumerate}\label{pro:basic properties lifts}
\item For every $v \in H$, $deg(v)=deg(\phi(v))$. In particular, if $G$ is $d$-regular then so is $H$.
\item If $f:V(G)\rightarrow\mathbb{R}$ in an eigenfunction of $G$ with eigenvalue $\lambda$,
then $f\circ\phi$ is an eigenfunction of $H$ with eigenvalue $\lambda$.
Such an eigenfunction-eigenvalue pair of $H$ is considered {\em old}. 
The other eigenfunctions and eigenvalues of $H$ are considered {\em new}.
\item If $G$ is disconnected then so is $H$.
\item $\chi(G)\ge\chi(H)$.
\item If $G$ is connected then all the fibers of vertices in $G$ have the same cardinality,
which we call the {\em covering number} of the lift.
\end{enumerate}
\end{proposition}

Let $G$ be fixed connected graph. Denote by $L_n(G)$
the collection of all lifts of $G$ with covering number $n$.
It is not hard to see that every
member $H \in L_n(G)$ has the following description.
It has vertex set $V(H)=V(G)\times[n]$ where the projection on the first
coordinate is the covering map from $H$ to $G$. To define the edges in $H$,
fix an arbitrary orientation on the edges of $G$ and associate a permutation $\sigma_e \in S_n$ to every edge $e\in E(G)$.
The edge set of $H$ is $E(H)=\{ (v,i)(u,\sigma_e(i)) : i\in[n],e=(v,u)\in E(G)\}$. This definition
lends itself naturally to randomization. In particular, $H$ is
a random $n$-lift if the $\sigma_e$'s are chosen uniformly at random from $S_n$.
It was shown in ~\cite{LP09} that for every $d$-regular $G$ it holds with probability
$1-o_n(1)$ that all new eigenvalues of a random $H \in L_n(G)$ are bounded by $O(d^{2/3})$.

\subsection{Tight products}
If $H$ is a lift of both $G_1$ and $G_2$, we say that $H$ is a {\em common lift} of these graphs.
This notion has been studied by Leighton~\cite{Lei82} who showed
that two finite graphs $G_1$ and $G_2$ have a common finite lift iff they share the same {\em universal cover}.
Thus, in particular, every two $d$-regular graphs have a common finite cover (as observed already by~\cite{AG81}).

In this paper we study a special kind of common lift.
\begin{definition}
A graph $H$ is called a {\em tight product} of graphs $G_1$ and $G_2$ if $V(H)=V(G_1)\times V(G_2)$ and 
both projection maps $V(H)\rightarrow V(G_1)$ and $V(H)\rightarrow V(G_2)$ are covering maps .
\end{definition}
This definition extends in the obvious way to tight products of more than two graphs.

In section \ref{se:existence and basic properties} we study some basic properties of tight products.
Specifically, we find conditions for its existence.
It turns out that $G_1$ and $G_2$ can have a tight product only if they are regular graphs 
of the same regularity. We also give sufficient conditions for the existence of a tight product.
E.g. for $d$ even every two $d$-regular graphs have a tight product.
On the other hand, some complication is to be expected here, because when $d$ is odd
it is $NP$-hard to determine whether a given pair of $d$-regular graphs has a tight product.

In section \ref{se:random models} we present some random models for regular graphs based on tight products.
We start with a $2d$-regular graph $G_B=G(\sigma_1,...,\sigma_d)$ as defined above,
where $\sigma_1,...,\sigma_d$ are permutations on $V(G_B)$.
We choose permutations $\pi_1,...,\pi_d \in S_n$ uniformly at random and define $H=G((\sigma_1,\pi_1),...,(\sigma_d,\pi_d))$.
Namely, $V(H)=V(G_B)\times[n]$ and $E(H)=\{(v,u)(\sigma_i(v),\pi_i(u)) :(v,u)\in V(H),i \in [d]\}$. 
Note that $H$ is a tight product of $G_B$ and the {\em random graph} $G_R=G(\pi_1,...,\pi_d)$. 
We use the trace method to show that all the new eigenvalues of $H$ are bounded (in absolute value) by $O(d^{3/4})$.
An adaptation of the methods of~\cite{LP09} might improve this upper bound to $O(d^{2/3})$.

An interesting feature of this model is that compared with the standard model of random lifts,
it offers a reduction in the necessary number of random bits. Whereas a random lift uses a
random permutation for each edge of the base graph, this model uses only $d$ random permutations. 
In addition, the generated graph has a concise representation. We discuss those aspects in the last 
subsection of section \ref{se:random models} and suggest some questions for future research.

\section{Existence and basic properties}\label{se:existence and basic properties}

\subsection{Basic properties}\label{se:basic properties}
In this section $H$ is always a tight product of $G_1$ and $G_2$. Here are some fairly obvious
consequences of proposition \ref{lifts basic properties} and the definition of tight product:

\begin{proposition}
\
\begin{enumerate}\label{pro:basic properties}
\item Every eigenvalue of $G_1$ or $G_2$ is also an eigenvalue of $H$
\item If $G_1$ or $G_2$ is bipartite then so is $H$.
\item If $G_1$ or $G_2$ is disconnected then so is $H$.
\item If both $G_1$ and $G_2$ are bipartite then $H$ is disconnected. 
\item If $G_1$ and $G_2$ have a tight product $H$, then $G_1$ and $G_2$ are $d$-regular with 
the same $d$.
\end{enumerate}
\end{proposition}
The only fact that needs some elaboration is 4. To see it,
let $B_1 \cup W_1$ and $B_2 \cup W_2$  be bipartitions of $V(G_1),V(G_2)$ respectively. There is no edge
between $(B_1\times W_2)\cup (B_2\times W_1)$ and $(B_1\times B_2)\cup (W_1\times W_2)$.\\

We now turn to prove that every pair of $2d$-regular graphs
has a tight product. This is a simple but useful observation.

\begin{proposition} \label{existence-basic}
\begin{enumerate}
\item Every two $2d$-regular graphs have a tight product. \label{2d-existence}
\item If both $G_1$ and $G_2$ are $(2d+1)$-regular and have a perfect matching,
then $G_1, G_2$ have a tight product. \label{2d+1-existence}
\end{enumerate}
\end{proposition}
\proof
Since $G_1$, $G_2$ are $2d$-regular graphs, they can be expressed as
$$G_1=G(\sigma_1,...,\sigma_d), G_2=G(\pi_1,...,\pi_d)$$
for some permutations $\sigma_i\in S_{V(G_1)},\pi_i\in S_{V(G_1)}$. We note that the graph
$$H=G((\sigma_1,\pi_1),...,(\sigma_d,\pi_d))$$
is a tight product of $G_1$, $G_2$. Indeed the neighbor set of the vertex
$(v,u)\in V(H)$ is $N_H((v,u))=\{(\sigma_i^{\pm 1}(v),\pi_i^{\pm 1}(u)):i\in [d]\}$.
It follows that the projection maps $V(H) \rightarrow V(G_1),H \rightarrow V(G_2)$ map $N_H((v,u))$ 
one to one and onto $N_{G_1}(v)$ and $N_{G_2}(u)$ respectively. 
The first claim follows.

The second claim can be proved in a similar manner, since a $(2d+1)$-regular
graph that contains a perfect matching is the union of $(d+1)$ permutations,
one of which is an involution with
no fixed points (and corresponds to the perfect matching). The proof follows along the same lines, but the edges that
correspond to the perfect matching are counted only once and not twice.
\proofbox

The above construction suggests a method for generating random
tight products. Start with a fixed $2d$-regular graph $G_1$ and 
pick a random $2d$-regular graph $G_2$. Now let $H$ be a tight product of $G_1$, $G_2$. 
To simplify matters, we can choose a fixed $2$-factorization of $G_1$ and compose $G_2$ from $d$ random permutations.
Alternatively, both $G_1$ and $G_2$ can be selected at random. In Section~\ref{se:random models} we investigate
the expansion properties of such graphs.

The rest of this section concerns
the problem of finding a tight product for a given pair of graphs.

\subsection{Class classifier and The computational hardness of finding a tight product}\label{hardness}
Proposition \ref{existence-basic} may suggest that every two $d$-regular graphs have a tight product. 
This is, however, not true as we observe below.\\
Recall Vizing's Theorem: If $\Delta(G)$ is the
largest vertex degree in $G$, then $G$'s edge-chromatic number, $\chi_E(G)$, is either $\Delta(G)$ or $\Delta(G)+1$.
Accordingly, $G$ is said to be of {\em class 1} or {\em class 2}.
It is known (\cite{Hol81}) that it is NP-Hard to determine the edge chromatic number, even if we restrict ourselves to cubic graphs. We prove that:
\begin{theorem}\label{theorem-class-classifier}
For every positive integer $k$, there is a $(2k+1)$-regular graph, $G^{2k+1}$, with the property that every $2k+1$-regular graph $G$ is of class-1 if and only if it has a tight product with $G^{2k+1}$. 
\end{theorem}
Consequently:
\begin{theorem}
The following decision problem TIGHTPRODUCT is NP-complete.\\
Input: Two finite graphs $G_1, G_2$.\\
Output: Do $G_1, G_2$ have a tight product?
\end{theorem}
Before we turn to prove theorem \ref{theorem-class-classifier} we discuss two notions - {\em neighborly permutations} and {\em semi-coloring}.

{\em Neighborly permutations:} Suppose that $H$ is a tight product of $G_1$ and $G_2$. As in every lift, every edge $\overrightarrow{v_1v_2}\in \vec E(G_1)$ defines a bijection $\sigma_{\overrightarrow{v_1v_2}}$ from the fiber of $v_1$ to the fiber of $v_2$ and we denote $\mathscr{P}_{G_1}(H)=\{\sigma_{\overrightarrow{v_1v_2}}:\overrightarrow{v_1v_2}\in \vec E(G_1)\}$. Since the lift $H$ is a tight product, $\sigma_{\overrightarrow{v_1v_2}}$ is a permutation of $V(G_2)$ that maps every vertex to one of its neighbors. A permutation $\sigma$ on the vertex set of a graph $G$ with this property is called {\em neighborly permutation}. Note that neighborly permutations correspond to oriented spanning subgraphs of $G$ that are the union of vertex-disjoint cycles (where we permit a single edge to be considered as a cycle as well).\\
Neighborly permutations are useful in characterizing tight products:

\begin{proposition}\label{pro:tight-product-def-via-neighborly-permutation}
Let $G_1$ and $G_2$ be $d$-regular graphs. Suppose that $H$ is a tight product of $G_1$ and $G_2$ with $\mathscr{P}_{G_1}(H)=\{\sigma_{\overrightarrow{v_1v_2}}:\overrightarrow{v_1v_2}\in \vec E(G_1)\}$. Then:
\begin{enumerate}
\item For every $\overrightarrow{v_1v_2}\in \vec E(G_1)$, $\sigma_{\overrightarrow{v_1v_2}}^{-1}=\sigma_{\overrightarrow{v_2v_1}}$.
\item For $v_1 \in V(G_1),u_1 \in V(G_2)$, the mapping $v\rightarrow \sigma_{\overrightarrow{v_1v}}(u_1)$ from $N_{G_1}(v_1)$ to $N_{G_2}(u_1)$ is one to one and onto.\label{condition-2}
\end{enumerate}
Conversely, consider a collection $\mathscr{P}=\{\sigma_{\overrightarrow{v_1v_2}}:\overrightarrow{v_1v_2}\in \vec E(G_1)\}$ of neighborly permutations of $G_2$, that satisfies the above conditions. There is a unique tight product $H$ of $G_1$ and $G_2$ with $\mathscr{P}_{G_1}(H)=\mathscr{P}$.
\end{proposition}
\proof
Suppose that $H$ is a tight product of $G_1,G_2$. As in every lift, each $\sigma_{\overrightarrow{v_1v_2}}\in\mathscr{P}_{G_1}(H)$ satisfies $\sigma_{\overrightarrow{v_1v_2}}^{-1}=\sigma_{\overrightarrow{v_2v_1}}$. For condition \ref{condition-2}, suppose that $\sigma_{\overrightarrow{v_1v_2}}(u)= \sigma_{\overrightarrow{v_1v_3}}(u)=w$ for $v_2,v_3 \in N_{G_1}(v_1)$. Then, $(v_1,u)(v_2,w),(v_1,u)(v_3,w)\in E(H)$. As a covering map, the projection $V(H)\rightarrow V(G_2)$, maps $N_H(v_1,u)$ one to one and onto $N_{G_2}(u)$, so we have $v_2=v_3$. We showed that the mapping is one to one and since $|N_{G_1}(v_1)|$ = $|N_{G_2}(u)| = d$ it is also onto.

Suppose now that $\mathscr{P}$ satisfies the above conditions. We define a tight product $H$ of $G_1$ and $G_2$ by setting $V(H)=V(G_1)\times V(G_2)$ and $E(H)=\{(v_1,u)(v_2,\sigma_{\overrightarrow{v_1v_2}}(u)):\overrightarrow{v_1v_2}\in \vec E(G_1), u\in V(G_2)\}$. It is clear that the projection $V(H)\rightarrow V(G_1)$ is a covering map and that $\mathscr{P}_{G_1}(H)=\mathscr{P}$. By condition \ref{condition-2}, for every $(v,u)\in E(H)$, the projection $\pi_2:V(H)\rightarrow V(G_2)$ maps $N_H(v,u)$ one to one and onto $N_{G_2}(u)$, so $\pi_2$ is a covering map and $H$ is indeed a tight product. The uniqueness of $H$ is clear.
\proofbox
\begin{note} Every regular graph $G$ has a neighborly permutation. To see that, consider the standard $2$-lift, $\hat{G}$ of $G$ (The vertex set is $\{1,2\}\times V(G)$ and $(i,v)$ is adjacent to $(j,u)$ iff $i\ne j$ and $vu\in E(G)$). The regular bipartite graph $\hat{G}$ contains a perfect matching $\hat{M}$ where the corresponding edges $M\subset E(G)$ form a collection of cycles (some of which may be single edges viewed as a cycle of length 2). We orient those cycles arbitrarily to obtain a neighborly permutation.
\end{note}

{\em Semi-coloring:} Let $G=(V,E)$ be a graph and let $\Delta=\Delta(G)$ be the largest vertex degree in $G$. A {\em semi-coloring} is a coloring of $E$ with color set $[\Delta]\cup {[\Delta]\choose 2}$, i.e. each color is either an element of $[\Delta]$ or an
unordered pair of such elements. In the latter case we view the edge as being colored ``half $i$ and half $j$''. The coloring must satisfy, for every $v\in V$:
\begin{enumerate}
\item For every $i\in [\Delta]$, the total weight of $i$ on the edges incident with $v$ is at most $1$.
\item For every $i\ne j\in [\Delta]$, there are either $0$ or $2$ edges colored $\{i,j\}$ incident with $v$.
\end{enumerate}
Note that, for $d$-regular graphs, the total weight of $i\in [d]$ on the edges incident with some vertex $v$ is exactly $1$. Also note that if $G$ is semi-colored then the collection of edges colored by a specific pair is a union of vertex-disjoint cycles.\\
This seemingly strange concept is related to tight products via the following proposition:
\begin{proposition}\label{pro:semi-coloring}
Let $G_1$,$G_2$ be $(2k+1)$-regular graphs such that $G_1$ is semi-colored and $G_2$ is of class-1. Then $G_1$ and $G_2$ have a tight product.
\end{proposition}
\proof
To prove the existence of a tight product of $G_1$ and $G_2$, we construct a collection $\mathscr{P}=\{\sigma_{\overrightarrow{v_1v_2}}:\overrightarrow{v_1v_2}\in \vec E(G_1)\}$ of neighborly permutations of $G_2$ that satisfies the conditions of proposition \ref{pro:tight-product-def-via-neighborly-permutation}.

Since $G_2$ is of class-1, there is 1-factorization $E(G_2)=\cup_{i=1}^{2k+1}F_i$. We define neighborly permutations on $G_2$ as follows: For $0\le i , j \le 2k+1$, $F_i\cup F_j$ is a union of vertex-disjoint cycles. Fix some orientation on those cycles and define $\pi_{ij}=\pi_{ji}$ to be the corresponding permutation. Note that $\pi_{ii}$ is an involution.\\
We now define:
\begin{enumerate}
\item If $v_1v_2\in E(G_1)$ is colored by $1\le i\le 2k+1$, we define $\sigma_{\overrightarrow{v_1v_2}}=\sigma_{\overrightarrow{v_2v_1}}=\pi_{ii}$.
\item The set of edges in $G_1$ colored by the pair $\{i,j\}$ is the union of vertex-disjoint cycles and we arbitrarily orient those cycles. If $v_1v_2\in E(G_1)$ is colored by $\{i,j\}$ and the orientation is $v_1\to v_2$, we define $\sigma_{\overrightarrow{v_1v_2}}=\pi_{i,j},\;\sigma_{\overrightarrow{v_2v_1}}=\pi^{-1}_{i,j}$.
\end{enumerate}
It is clear that $\mathscr{P}$ satisfies the first condition in proposition \ref{pro:tight-product-def-via-neighborly-permutation}.
To see that condition 2 holds, we note that if $v_1v_2,v_1v_3\in E(G_1)$ with $v_2\ne v_3$ then 
$\sigma_{\overrightarrow{v_1v_2}}(u_1)\ne\sigma_{\overrightarrow{v_1v_3}}(u_1)$ for every $u_1\in V(G_2)$. 
For example, if $v_1v_2$ is colored by $i\in [d]$, and $v_1v_3$ is colored by $j\ne i$, then 
$\sigma_{\overrightarrow{v_1v_2}}(u_1)$ is the vertex that is matched to $u_1$ by the matching $F_i$ and 
$\sigma_{\overrightarrow{v_1v_3}}(u_1)$ is the vertex that is matched to 
$u_1$ by the matching $F_j$. Since $F_i$ and $F_j$ are disjoint, those vertices are different. 
Consequently, the mapping $v\rightarrow\sigma_{\overrightarrow{v_1v}}(u_1)$ is one to one, 
and since both graphs are $d$-regular, it is also onto and condition 2 holds.  
\proofbox

\proof (of Theorem \ref{theorem-class-classifier}).
We postpone the construction of $G^{2k+1}$ to the end of the proof, and mention two properties it has on which we rely:
\begin{enumerate}
\item There is a vertex, $v_0 \in V(G^{2k+1})$ that does not belong
to any proper cycle, i.e. all edges incident with $v_0$ are bridges.
\item $G^{2k+1}$ has a semi-coloring.
\end{enumerate}
By proposition \ref{pro:semi-coloring}, every $(2k+1)$-regular graph of class-1 has a tight product with $G^{2k+1}$. Conversely let $G$ be a $(2k+1)$-regular graph, and suppose that $H$ is a tight product of $G$ and $G^{2k+1}$. We must show that $G$ is of class-1.

Denote $\mathscr{P}_G(H)=\{\sigma_{\overrightarrow{v_1v_2}}:\overrightarrow{v_1v_2}\in \vec E(G)\}$.
First, we claim that $\sigma_{\overrightarrow{v_1v_2}}(v_0)=\sigma_{\overrightarrow{v_2v_1}}(v_0)$ for every $\overrightarrow{v_1v_2}\in \vec E(G)$. To this end,
express the permutation $\sigma_{\overrightarrow{v_1v_2}}=(u_{11},u_{12},\ldots,u_{1\nu_1}) \cdots (u_{r1},u_{r2},\ldots,u_{r\nu_r})$ as a product of disjoint cycles. By the defining property of neighborly permutation, 
for all indices $t$, $(u_{t1},u_{t2},\ldots,u_{t\nu_t})$ is a
(graph theoretic) simple cycle in $G^{2k+1}$. But the only simple cycles in $G^{2k+1}$ that contain the vertex $v_0$ are of length 2 (i.e. a single edge), hence, $\sigma_{\overrightarrow{v_1v_2}}(v_0)=\sigma^{-1}_{\overrightarrow{v_1v_2}}(v_0)=\sigma_{\overrightarrow{v_2v_1}}(v_0)$ as required.

By the last discussion, we can define a $(2k+1)$-edge coloring $c$ of $E(G)$ as follows: $c(v_1v_2)=\sigma_{\overrightarrow{v_1v_2}}(v_0)$. To see that this yields a proper edge coloring of $G$, consider two incident edges $uv_1, uv_2 \in E(G),v_1\ne v_2$. By proposition \ref{pro:tight-product-def-via-neighborly-permutation}, $c(uv_1)=\sigma_{\overrightarrow{uv_1}}(v_0)\ne \sigma_{\overrightarrow{uv_2}}(v_0)=c(uv_2)$.

{\em The construction of $G^{2k+1}$}. We take $k$ copies of $K_{2k+2}$ and remove one edge from each copy. We add new vertex called the {\em secondary pivot} and connect it to every vertex that belonged to one of the removed edges. The graph we obtained is called {\em cluster}. To construct $G^{2k+1}$, we start with $2k+1$ clusters and add a new vertex called the {\em main pivot} and connect it to each of the secondary pivots. We enumerate the pivots by $\{0,\ldots ,2k+1\}$ where $0$ is the main pivot and $1,\ldots ,2k+1$ are the secondary pivots. A picture is worth more than thousand words.

\begin{figure}[ht]
\centering
\input{tight_product_pic_2.epic}
\caption{$G^5$}
\label{fig:bla}
\end{figure}

It is clear that no cycle goes through the vertex $0$, so it only remains to find a semi-coloring of $G^{2k+1}$.\\
For $1\le i \le 2k+1$, consider the subgraph $H_i$ of $G^{2k+1}$ that is the cluster whose secondary pivot is $i$ together with the edge $i0$. We can choose a perfect matching $M_i\subset E(H_i)$ such that $i0\in M_i$. The graph that is obtained from $H_i$ upon removal of the edges in $M_i$ and the vertex $0$ is $2k$-regular. Therefore, it has $2$-factorization $F_i^1,...,F_i^k$. We decompose $[2k+1]\setminus \{i\}$ into $k$  pairs $c_1,...,c_k$. For each $v_1v_2\in E(H_i)$, if $v_1v_2\in M_i$ we color $v_1v_2$ by $i$, and if $v_1v_2 \in F_i^j$ we color $v_1v_2$ by $c_j$.
It easy to check that this coloring is semi-coloring.

\proofbox

\subsection{Existence of tight products for cubic graphs and Vizing's theorem}\label{odd regularity}
This section is devoted to the following claim:
\begin{theorem}\label{theorem:3-clorable-graph-has-a-tight-product-with-every-cubic-graph}
Every graph $G$, with $\Delta(G)=3$ has a semi-coloring. Consequently, a class-1 cubic graph has a tight product with {\em every} cubic graph.
\end{theorem}
As an aside, we obtain a new proof to Vizing's theorem for the case of cubic graphs. 
\proof
It is convenient to introduce some terminology for the available colors - we denote Blue=1, Red=2, Green=3, Bright Blue=$\{ 2 , 3 \}$, Bright Red=$\{1,3\}$, Bright Green=$\{1,2\}$.\\
Let us assume first that $G$ is bridgeless, in particular, all vertices have degree $2$ or $3$.\\
If $G$ is cubic, then by Petersen's theorem, it has a perfect matching $M\subset E(G)$. We color all the edges in $M$ blue and the rest of the edges bright blue. This is a semi-coloring.\\
Suppose now that $\{v_1,\ldots ,v_k\}$ is
the set of degree 2 vertices in $G$ and $k\ge 1$. We can construct a $3$-regular graph $\hat{G}$, with at most one bridge, that contains a copy of $G$ as an induced subgraph of $\hat{G}$ (E.g. take two copies of $G$ and connect each corresponding pair of
degree 2 vertices). By Petersen's theorem, $\hat{G}$ has a perfect matching and like in the previous discussion, $\hat{G}$ has a semi-coloring $\hat{\phi}$ (note that we used a version of Petersen's theorem claiming that every cubic graph with {\em at most two} bridges have a perfect matching). Let $\phi$ be the restriction of $\hat{\phi}$ to $G$. Clearly, the conditions for semi-coloring  hold for $\phi$ at every vertex $v$ of degree 3. It might happen, however, that a degree-2 vertex $v$ has exactly one brightly colored edge, say bright blue. However, in this case $v$ is the end vertex of a bright blue path $P$. We recolor $P$ alternately red and green instead, thus eliminating the problem without creating any new problematic vertices. Repeating this procedure, if necessary, concludes with a semi-coloring.

If $G$ contains a bridge $e$ we remove it and deal separately with the two components using induction. By renaming the colors, if necessary, at one of the two components, we can combine them and color $e$ as well to semi-color the whole $G$.

Although this proof is formulated in the language of simple graphs, it carries through easily also when $G$ may include parallel edges or
loops. The only thing worth mentioning is that it is easy to observe that Petersen's Theorem remains valid in this case.
\proofbox

Our approach sheds some new light on Vizing's classical theorem.

\begin{theorem}\label{theorem:vizing}
[Vizing's theorem for cubic graphs] Every cubic graph $G$ can be 4-edge-colored.
\end{theorem}
\proof 
We start with a semi-coloring of $G$. Let $C$ be a cycle colored brightly, say bright-red. If $C$ has even length we recolor its edges green/blue alternately. If $C$ has odd length, we do likewise, except that the last edge is given our fourth color.
\proofbox
 
Finally we derive a necessary condition for two cubic graphs to have a tight product.
\begin{proposition}
Let $H$ be a tight product of the graphs $G_1,G_2$. If $G_2$ has a bridge then $G_1$ has a perfect matching.
\end{proposition}
\proof
Suppose $u_1u_2\in E(G_2)$ is a bridge. Denote $\mathscr{P}_{G_1}(H)=\{\sigma_{\overrightarrow{v_1v_2}}:\overrightarrow{v_1v_2}\in \vec E(G_1)\}$ and define $M=\{v_1v_2\in E(G_1) :  \sigma_{\overrightarrow{v_1v_2}}(u_1)=u_2\}$. Since $u_1u_2$ is a bridge, $M$ is well defined - i.e. $\sigma_{\overrightarrow{v_1v_2}}(u_1)=u_2\Leftrightarrow \sigma_{\overrightarrow{v_2v_1}}(u_1)=u_2$ (As in the proof of Theorem \ref{theorem-class-classifier}).

By proposition \ref{pro:tight-product-def-via-neighborly-permutation}, for every $v_1\in V(G_1)$ there is exactly one $v_2\in N(v_1)$ such that $\sigma_{\overrightarrow{v_1v_2}}(u_1)=u_2$, so $M$ is a perfect matching.
\proofbox
\begin{corollary}
If two cubic graphs have a tight product, at least one of them has a perfect matching.
\end{corollary}
\proof
Suppose that $G_1,G_2$ are cubic graphs having a tight product. If $G_2$ is bridgeless, it has a perfect matching by Petersen Theorem. Otherwise, the above proposition implies that $G_1$ has a perfect matching.
\proofbox
\subsection{Conclusion and open questions}\label{se:open question}
Let $G_1,G_2$ be $d$-regular graphs with $d\ge 3$ odd. Table \ref{table} summarizes our knowledge and open questions regarding the existence of a tight product of $G_1$ and $G_2$.
\\\begin{figure}[ht]
\centering

\begin{tabular}{ |p{3cm}|p{3cm}|p{3cm}|p{3cm} }
\hline
\backslashbox{$G_1$ is}{$G_2$ is}  & class-1 & class-2 with a perfect matching & \multicolumn{1}{|p{3cm}|}{class-2 without a perfect matching} \\
\hline
class-1 & always exists \\
\cline{1-3}
class-2 with a perfect matching& always exists& \multicolumn{1}{|c|}{always exists}\\
\cline{1-4}
class-2 without a perfect matching& always exists for $d=3$ (Open question for \mbox{$d>3$})  & May not exist (We do not know a case where it exists)& \multicolumn{1}{|p{3cm}|}{Does not exist for $d=3$ (Open question for \mbox{$d>3$})}\\
\cline{1-4}
\end{tabular}
\caption{When does a tight product exist for two $d$-regular graphs ($d$ odd)?}
\label{table}
\end{figure}


Completion of the bottom left box in the table might be achieved by answering the following question: 
\begin{open question}
Does every graph have a semi-coloring?
\end{open question}
In this context, we note that the following families of graphs can be semi-colored:
\begin{enumerate}
\item Class-1 graphs (The $\Delta(G)$-edge-coloring will do).
\item $2k$-regular graphs (Find a 2-factorization and color the $i$-th factor half $i$ half $k+i$).
\item $(2k+1)$-regular graphs containing a perfect matching (Use one color for the perfect matching. The remaining graph is $2k$-regular and can be handled as above).
\item Graphs with maximum degree $\le 3$ (Theorem \ref{theorem:3-clorable-graph-has-a-tight-product-with-every-cubic-graph}).
\end{enumerate}
It is of interest as well to seek tight products with additional properties, e.g.,:
\begin{open question}
Assume that $G_1,G_2$ have a tight product. When do they have a connected tight product?
What can be said about the possible chromatic number of their tight products. Likewise for
other graph parameters.
\end{open question}
We also wonder whether Vizing's theorem can be proved in full along the same lines of theorem \ref{theorem:vizing}.

\section{Random models}\label{se:random models}

The spectrum of graph lifts (and more specifically random lifts of
of graphs) has attracted considerable interest. We now consider some basic questions in this vein
in the context of tight products.

\subsection{Background: Word maps}\label{Background:-Word-maps}
Let $\Sigma_d$ be the alphabet consisting of the letters $g_1^{\pm 1},\ldots,g_d^{\pm 1}$. We denote by $\Sigma_d^k$ the set of all the words $\omega=g_{i_1}^{j_1}\cdot\cdot\cdot g_{i_{k}}^{j_{k}}$ with $j_{\nu} \in \{-1,1\}$
of length $k$ with letters from $\Sigma_d$. We associate a {\em word map} $\overline{\omega} : S_n^k\rightarrow S_n$
with every word $\omega=g_{i_1}^{j_1}\cdot\cdot\cdot g_{i_{k}}^{j_{k}} \in \Sigma_d^k$ as follows:
For every $(\sigma_1,...,\sigma_d) \in S_n^d$ we define $\overline{\omega} (\sigma_1,...,\sigma_d)=\sigma_{i_k}^{j_k}\circ...\circ \sigma_{i_1}^{j_1}$. With the uniform probability measure on $S_n^k$, $\overline{\omega}$ is a random (not necessarily uniform) permutation. We are interested in fixed points of such permutations and define $p(\omega)=Pr[\overline{\omega}(1)=1]$.

A word $\omega \in \Sigma^k$ is
called {\em reduced} if it does not contain two inverse consecutive letters. If $\omega \in \Sigma^k$ is not reduced, we can drop a pair of consecutive inverse letters. This can be repeated until a reduced word is attained.  We denote the resulting word by $\omega'$. It is not hard to see that $\omega '$ does not depend on the order at which reduction steps are performed. It is clear that $\overline{\omega}=\overline{\omega}'$, so $p(\omega)=p(\omega')$. We now define the {\em order} of $\omega$, denoted by $o(\omega)$, to be the largest integer $l$, such that $\omega'$ can be written as $\omega'=\omega_a \omega_b^l (\omega_a)^{-1}$ with nonempty $\omega_b$ (and $o(\omega)=0$ when $\omega'$ is empty). If $o(\omega)=1$ we say that $\omega$ is {\em primitive}.

Bounds on $p(\omega)$ are the backbone of the analysis we'll present in the next subsection as well as in many theorems regarding expansion of random graphs (e.g. \cite{Fri03}). We now state two such bounds (proofs can be found in \cite{HLW06}). The first lemma says that
for a primitive $\omega$ the behavior of $\omega (\sigma_1,...,\sigma_k)$ resembles
that of a random permutation. The second lemma bounds the number of imprimitive words.
\begin{lemma}
Let $\omega \in \Sigma^k$ be a primitive word. Then $p(\omega)\le \frac{1}{n-k}+\frac{k^4}{(n-k)^2}$.
\end{lemma}
\begin{lemma}
The number of imprimitive words in $\Sigma^{2k}$ is  $\le k^2(2\sqrt{2d})^{2k}$
\end{lemma}
A refined analysis of word maps can be found in \cite{LP09}. The (more involved) method of that paper might yield better bounds than what's shown below.

\subsection{First model - fixed base graph}
Fix a positive integer $n$ and a $2d$-regular graph $G_B$ expressed as
$$G_B=G(\sigma_1,...,\sigma_d)$$
where $\sigma_1,...,\sigma_d$ are permutation on $V(G_B)$.
Choose permutations $\pi_1,...,\pi_d \in S_n$ uniformly and independently at random and define:
$$H=G((\sigma_1,\pi_1),...,(\sigma_d,\pi_d))$$
$$G_R=G(\pi_1,...,\pi_d)$$
$H$ is called the {\em random product} of
the {\em random graph} $G_R$ with the {\em base graph} $G_B$.
Note that $H$ is indeed a tight product of $G_B$ and $G_R$ (see proposition \ref{existence-basic}).
By proposition \ref{pro:basic properties}, all the eigenvalues of $G_B$ (as well as of $G_R$) are also eigenvalues of $H$. We use the trace method to bound $\mu(H)$ - the absolute value of the largest new eigenvalue of $H$.\\
In \cite{Fri03} Friedman proved that the largest new eigenvalue in a (standard) lift of $2d$-regular graph is bounded in absolute value by $O(d^{\frac{3}{4}})$ a.a.s. The proof we present is an adaptation of his proof.

\begin{theorem}
$$E[\mu(H)]\le\ 32^{1/4}\cdot d^{3/4}+o(1)$$
Consequently, $\mu(H)=O(d^{\frac{3}{4}})$ a.a.s. as $n\to\infty$.
\end{theorem}
\proof
Denote by $A_H,A_{G_B}$ the adjacency matrices of $H$ and $G_B$. By Jensen's inequality,
\begin{equation}\label{mu bound}
(E[\mu(H)])^{2k} \le E[\mu(H)^{2k}] \le E[\sum\ \lambda^{2k}] =E[Tr(A_H^{2k})]-Tr(A_{G_B}^{2k})
\end{equation}
Where the sum in the third expression is over
all new eigenvalues. But $Tr[A_H^{2k}]$ has a combinatorial interpretation - it counts the {\em closed paths} of length $2k$ in $H$.

Denote by $P_{G_B}^k$ the set of all paths of length $k$ in $G_B$. We view $P_{G_B}^k$ as the set $V(G_B)\times \Sigma_d^k$ in the following manner: given a pair $(v_0,g_{i_1}^{j_1}\cdot\cdot\cdot g_{i_k}^{j_k})$, the corresponding path is $v_0\rightarrow v_1\rightarrow\cdot\cdot\cdot\rightarrow v_k$ where $v_t=\sigma_{i_t}^{j_t}(v_{t-1})$. It is clear that this correspondence is a bijection between $V(G_B)\times \Sigma^k$ and $P_{G_B}^k$. In the same manner, we denote the paths of length $k$ in $H$ and in $G_R$ by $P_H^k$ , $P_{G_R}^k$ respectively (and associate them with $V(H)\times \Sigma_d^k,V(G_R)\times \Sigma_d^k$). We denote by $C_{G_B}^k \subset P_{G_B}^k $ the set of closed paths in $G_B$.

Given $(u,\omega) \in P_{G_R}^k$, denote by $1_{G_R}^{(u,\omega)}$ the indicator function of the event that $(u,\omega)$ is a closed path in $G_R$ and observe that $E[1_{G_R}^{(u,\omega)}]=p(\omega)$. We define $1_{G_B}^{(v,\omega)}$,$1_{H}^{((v,u),\omega)}$ similarly. It is clear that a path $((v,u),\omega)$ in $H$ is closed iff its projections on $G_B$ and $G_R$ are both closed. Consequently, \mbox{$1_H^{((v,u),\omega)}=1_{G_B}^{(v,\omega)} \cdot 1_{G_R}^{(u,\omega)}$.}
With these notations and the lemmas from the previous subsection we obtain:
\begin{eqnarray}
E[Tr(A_H^{2k})] &=& E[\sum_{((v,u),\omega)\in P_H^{2k}}\ 1_H^{((v,u),\omega)}]\notag\\
&=& E[\sum_{((v,u),\omega)\in P_H^{2k}}\ 1_{G_B}^{(v,\omega)} \cdot 1_{G_R}^{(u,\omega)}]\notag\\
&=& E[\sum_{(v,\omega) \in P_{G_B}^{2k}}\ 1_{G_B}^{(v,\omega)} \sum_{u \in V(G_R)}\ 1_{G_R}^{(u,\omega)}]\notag\\
&=& \sum_{(v,\omega) \in C_{G_B}^{2k}}\ \sum_{u \in V(G_R)}\ E[1_{G_R}^{(u,\omega)}]\notag\\
&=& \sum_{(v,\omega) \in C_{G_B}^{2k}}\ n \cdot p(\omega)\notag
\end{eqnarray}
Following \cite{Fri03}, we split the sum according to whether $\omega$ is primitive or not. For non-primitive $\omega$, we overestimate $p(\omega)$ by $1$. We then use the lemmas from subsection (\ref{Background:-Word-maps}).
\begin{eqnarray*}
&& = \sum_{(v,\omega) \in C_{G_B}^{2k},o(\omega)=1}\ n \cdot p(\omega)+\sum_{(v,\omega) \in C_{G_B}^{2k},o(\omega)\ne1}\ n \cdot p(\omega)\notag\\
&&\le[\sum_{(v,\omega) \in C_{G_B}^{2k},o(\omega)=1}\ n \cdot p(\omega)]+n|V(G_B)||\{\omega : o(\omega)\ne 1\}|\notag\\
&&\le[\sum_{(v,\omega) \in C_{G_B}^{2k},o(\omega)=1}\ \frac{n}{n-2k}+\frac{n\cdot(2k)^4}{(n-2k)^2}]+n|V(G_B)|k^2(2\sqrt{2d})^{2k}\notag\\
&&\le Tr[A_G^{2k}]+[\sum_{(v,\omega) \in C_{G_B}^{2k},o(\omega)=1}\ \frac{2k}{n-2k}+\frac{n\cdot(2k)^4}{(n-2k)^2}]+n|V(G_B)|k^2(2\sqrt{2d})^{2k}\notag
\end{eqnarray*}

By (\ref{mu bound}) we obtain:
\begin{eqnarray*}
(E[\mu(H)])^{2k} &\le& [\sum_{(v,\omega) \in C_{G_B}^{2k},o(\omega)=1}\ \frac{2k}{n-2k}+\frac{n\cdot(2k)^4}{(n-2k)^2}]+n|V(G_B)|k^2(2\sqrt{2d})^{2k}\notag\\
\end{eqnarray*}
For $k=o(n)$ (as we actually assume below),
\begin{eqnarray}
(E[\mu(H)])^{2k} &\le& Tr[A_{G_B}^{2k}]\frac{5(2k)^4}{n}+n|V(G_B)|k^2(2\sqrt{2d})^{2k}\notag\\
&\le& |V(G_B)|(2d)^{2k}\cdot\frac{5(2k)^4}{n}+n|V(G_B)|k^2(2\sqrt{2d})^{2k}\label{trace bound 3}
\end{eqnarray}
The last inequality is justified since every entry on the diagonal of $A_{G_B}$ is bounded by $(2d)^{2k}$.
To finish, we choose $k=\log_{\frac{2d}{2\sqrt{2d}}}(n)\iff n=(\frac{2d}{2\sqrt{2d}})^k$. Then, from (\ref{trace bound 3}) we obtain:
\begin{eqnarray*}
(E[\mu(H)])^{2k} &\le& |V(G_B)|5(2k)^4(4d\sqrt{2d})^k+|V(G_B)|k^2(4d\sqrt{2d})^k\\
&\le& |V(G_B)|100k^4(4d\sqrt{2d})^k
\end{eqnarray*}
Therefore,
$$E[\mu(H)] \le (|V(G_B)|100k^4)^{\frac{1}{2k}}\sqrt{4d\sqrt{2d}}=\sqrt{4d\sqrt{2d}}+o(1)$$
\proofbox

\subsection{Conclusions and an open problem}
How large can $\mu(H)$ be? Again, following \cite{Fri03} we raise:
\begin{conjecture}
Let $H$ be a random tight product as defined in the beginning of subsection \ref{Background:-Word-maps}. Then, for every $\epsilon > 0$, $\mu(H) \le 2\sqrt{2d-1}+\epsilon$ a.a.s.
\end{conjecture}
The potential advantage of this conjecture over Friedman's, is that it may allow to construct graphs with a near optimal spectral gap, using very limited randomness. More generally it is of interest to
find a distribution $\mu$ on $S_n$, with {\em small entropy}, such that if we choose $\sigma_1,...,\sigma_d$ independently at random from the distribution $\mu$, then $G(\sigma_1,...,\sigma_d)$ has small second eigenvalue w.h.p.
In this context we should mention
\cite{BG08}, where it is shown that there is $\kappa < 2d$ such that if 
$g_1,\ldots,g_d\in SL_2(\mathbb F_p)$ are chosen uniformly and independently at random, then the spectral radius of the Cayley graph of $SL_2(\mathbb F_p)$
with generates $\{g_1,g_1^{-1}\ldots,g_d,g_d^{-1}\}$ is a.a.s. (with respect to $p$) bounded by $\kappa$.

Tight products suggest another approach to this problem, as follows:
Consider $S_n^k$ as a subset of $S_{n^k}$. Here we allow $k$ to grow with $n$,
so that $|S_n^k|$ is much smaller than $|S_{n^k}|$ and we indeed save in entropy.
Does this yield an expander family? Can we do this even with $k$ that grows with $n$?

\section*{Acknowledgements}
We would like to thank Alex Lubotzky and Baoyindureng Wu for insightful comments.

\end{document}